\documentclass[a4paper]{article}

\usepackage{INTERSPEECH2021}
\usepackage{multirow}
\usepackage{xcolor}

\title{Relaxing the Conditional Independence Assumption of CTC-based ASR \\ by Conditioning on Intermediate Predictions}
\name{Jumon Nozaki$^{1, 2}$, Tatsuya Komatsu$^1$}
\address{
  $^1$LINE Corporation, Japan\\
  $^2$Kyoto University, Japan}
\email{\{nozaki.jumon, komatsu.tatsuya\}@linecorp.com}

\begin{document}

\maketitle

\begin{abstract}
% The total length of the abstract is limited to 200 words. 
This paper proposes a method to relax the conditional independence assumption of connectionist temporal classification (CTC)-based automatic speech recognition (ASR) models.
We train a CTC-based ASR model with auxiliary CTC losses in intermediate layers in addition to the original CTC loss in the last layer.
During both training and inference, each generated prediction in the intermediate layers is summed to the input of the next layer to condition the prediction of the last layer on those intermediate predictions.
Our method is easy to implement and retains the merits of CTC-based ASR: a simple model architecture and fast decoding speed.
We conduct experiments on three different ASR corpora.
Our proposed method improves a standard CTC model significantly (e.g., more than 20 \% relative word error rate reduction on the WSJ corpus) with a little computational overhead.
Moreover, for the TEDLIUM2 corpus and the AISHELL-1 corpus, it achieves a comparable performance to a strong autoregressive model with beam search, but the decoding speed is at least 30 times faster.
%  167/200 words

\end{abstract}

\section{Introduction}

Fueled by the drastic development of deep neural networks, End-to-end (E2E) automatic speech recognition (ASR) systems have achieved promising performance~\cite{bahdanau2016end, chiu2018state, karita2019comparative} and become mainstream in scientific research of ASR.
There are two main types of E2E models: attention-based encoder-decoder (AED) models~\cite{bahdanau2016end, dong2018speech} and Connectionist Temporal Classification~\cite{graves2006connectionist} (CTC) models~\cite{graves2014towards, kriman2020quartznet}.
The AED model is a model consisting of an encoder, which encodes acoustic features, and a decoder, which generates a sentence.
The architectures of many state-of-the-art ASR systems~\cite{karita2019improving, gulati2020conformer} are based on the AED models.
However, the AED model outputs tokens one by one in an autoregressive manner, where each token depends on all previously generated tokens, resulting in recognition latency.
On the other hand, the CTC model consists of only an encoder and outputs all tokens independently in a non-autoregressive manner.
Although its decoding speed is faster than the one of the AED model, it is generally inferior in terms of recognition accuracy due to the conditional independence assumption between output tokens.
In recent years, there has been a demand for technologies to run ASR systems on devices such as smartphones and tablets.
Those devices typically have only limited computing resources, thus a fast and light inference process is an important factor in on-device ASR.
Therefore, improving the accuracy of the CTC model is an important research area.

There also has been a lot of work done exploring non-autoregressive speech recognition models other than the aforementioned standard CTC model~\cite{chen2019listen, higuchi2020mask, fujita2020insertion}, inspired by work in the area of neural machine translation~\cite{gu2017non, ghazvininejad2019mask, stern2019insertion}.
One popular approach is to refine an output by iterative decoding to relax the conditional independence assumption between tokens.
Chen et al.~\cite{chen2019listen} proposed a model based on the conditional masked language model~\cite{ghazvininejad2019mask}, in which mask tokens fed to the decoder are iteratively predicted.
Several papers proposed to leverage the output of CTC as the initial input of the refinement process~\cite{higuchi2020mask, chi2020align, higuchi2020improved}.
Mask-CTC~\cite{higuchi2020mask} inputs the greedy CTC output to the decoder and refines low-confidence tokens conditioning on high-confidence tokens, while Align-Refine~\cite{chi2020align} inputs a latent alignment of CTC to the decoder and refines over the alignment space.
However, these models need complicated model structures and training methods compared to the standard CTC model.
Also, they sacrifice inference speed by incorporating iterative decoding.

Apart from these aforementioned papers, a model called InterCTC~\cite{lee2021intermediate} was proposed to improve the accuracy of CTC itself using neither a decoder nor iterative decoding.
InterCTC is trained with intermediate CTC loss, which is calculated over the intermediate
representation of the model, in addition to the original CTC loss.
InterCTC achieves a performance improvement compared to the standard CTC model with no computational overhead during inference.
Because InterCTC is trained with intermediate CTC loss, we assume it can already recognize speech in intermediate layers to some extent.
We call probability distributions over vocabularies, which are calculated to take CTC loss in intermediate layers, intermediate predictions, and assume they can be utilized to further improve accuracy.

In this work, we propose to make use of intermediate predictions to relax the conditional independence assumption of CTC-based ASR models by conditioning the final prediction on the intermediate predictions.
Specifically, in both training and inference time, we add the intermediate prediction to the input of the next encoder layer in order for hidden embeddings to include the information of predictions of previous layers.
Our method can be regarded as performing refinement such as Mask-CTC and Align-Refine inside the encoder without using a decoder and without requiring a complex training method.
Our method not only improves recognition accuracy with a little degradation of inference speed of CTC but also is easy to implement as it only requires a few modules to be added to the existing models that have stacked encoders.

\begin{figure*}[t]
\centering
\includegraphics[width=\linewidth]{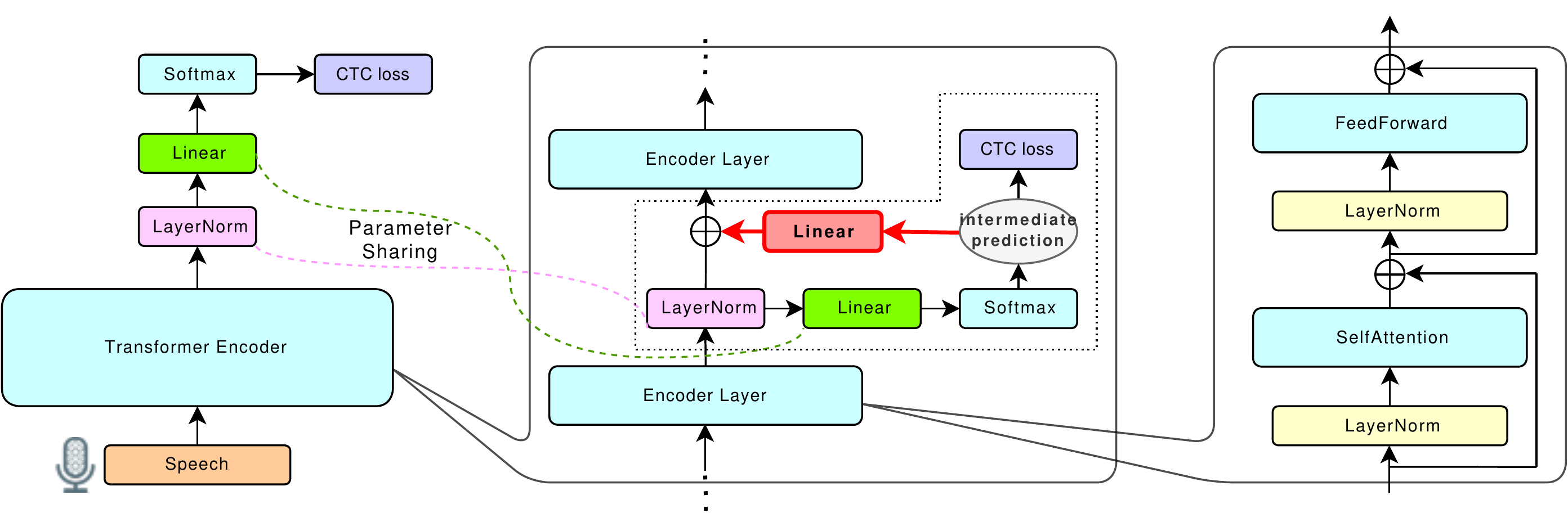}
\caption{Illustration of our proposed method.
We take intermediate CTC loss for outputs of some encoder layers in addition to the CTC loss for the output of the final encoder layer.
The intermediate prediction is summed to the input of the next encoder layer to condition the prediction of the final layer on it.
The parameters of the red linear layer for transforming the dimension of the intermediate prediction are shared in the model.
The area enclosed with a dotted line between encoder layers only exists after some encoder layers.
}
\label{fig:proposed_model}
\end{figure*}

\section{Methods}
\subsection{Connectionist Temporal Classification}
We let $\mathbf{X} \in \mathbb{R}^{S \times D}$ be a $D$-dimensional input sequence of length $S$ and $\mathbf{y} \in \mathbb{\mathcal{V}}^{T}$ be an output sequence of length $T$, where $\mathcal{V}$ is a vocabulary.
% Each element $y_i \in \mathcal{V}$ is a token in the output sequence $\mathbf{y}$.
CTC~\cite{graves2006connectionist} is trained to maximize the log-likelihood over all valid alignments $\mathbf{a} \in \mathbb{\mathcal{V}^{\prime}}^{S}$ between $\mathbf{X}$ and $\mathbf{y}$.
Here, $\mathcal{V}^{\prime} = \mathcal{V} \cup \{\epsilon\}$, where $\epsilon$ is a special blank token.
The log-likelihood is defined as follows:
\begin{equation}
   \log P_{\mathrm{CTC}}(\mathbf{y} \mid \mathbf{X})=\log \sum_{\mathbf{a} \in \Gamma^{-1}(\mathbf{y})} P(\mathbf{a} \mid \mathbf{X})
\end{equation}
where $\Gamma$ is the function of $\mathbf{a}$, which removes all special blank tokens and repeated labels.
The probability of each alignment $\mathbf{a}$ is modeled with the conditional independence assumption between tokens:
\begin{equation}
    P(\mathbf{a} \mid \mathbf{X})= \prod_{s} P(a_{s} \mid \mathbf{X})
\end{equation}
where $a_s$ denotes the $s$-th symbol of $\mathbf{a}$.
Thanks to the conditional independence assumption, CTC can generate tokens in parallel thus achieving fast decoding speed, but cannot model language dependencies between tokens.

\subsection{Transformer Encoder for CTC-based ASR}
We use the encoder of Transformer~\cite{vaswani2017attention} consisting of $L$ encoder layers.
The $l$-th encoder layer transforms its input $\mathbf{X}_{l}^{\mathrm{in}} \in \mathbb{R}^{S \times D}$ into $\mathbf{X}_{l}^{\mathrm{out}} \in \mathbb{R}^{S \times D}$ as follows:
\begin{equation}
    \mathbf{X}_{l}^{\prime} =\text { SelfAttention }(\mathbf{X}_{l}^{\mathrm{in}})+\mathbf{X}_{l}^{\mathrm{in}}
\end{equation}
\begin{equation}
    \mathbf{X}_{l}^{\mathrm{out}} =\mathrm{FFN}(\mathbf{X}_{l}^{\prime})+\mathbf{X}_{l}^{\prime}
\end{equation}
where $\text{ SelfAttention }\left( \cdot \right)$ and $\text{ FFN }\left( \cdot \right)$ denote the self-attention layer and the feedforward network with layer normalization~\cite{ba2016layer}.
Then, the output of the $l$-th encoder layer is directly used as the input of the $(l+1)$-th encoder layer:
\begin{equation}
    \mathbf{X}_{l+1}^{\mathrm{in}} = \mathbf{X}_{l}^{\mathrm{out}}
\end{equation}

To train a Transformer encoder for ASR with CTC loss criterion, the output of the last encoder layer $\mathbf{X}_{L}^{\mathrm{out}}$ is first mapped to a posterior probability distribution over vocabularies $\mathbf{Z}_{L} \in \mathbb{R}^{S \times \mathcal{|V^{\prime}|}}$ using a layer normalization layer, a linear layer and softmax function:
\begin{equation}
    \mathbf{X}_{L}^{\mathrm{norm}} = \mathrm{LayerNorm}(\mathbf{X}_{L}^{\mathrm{out}})
\end{equation}
\begin{equation}
    \mathbf{Z}_{L} = \mathrm{Softmax}(\mathrm{Linear}_{D \rightarrow \mathcal{|V^{\prime}|}}(\mathbf{X}_{L}^{\mathrm{norm}}))
\end{equation}
where $\mathrm{LayerNorm}(\cdot)$, $\mathrm{Softmax}(\cdot)$, and $\mathrm{Linear}_{D \rightarrow \mathcal{|V^{\prime}|}}(\cdot)$ denote a layer normalization layer, a softmax function, and a linear layer that maps vectors of the dimension $D$ to ones of the dimension $\mathcal{|V^{\prime}|}$.
Then, the log probability of Eq. (1) is calculated as an objective function using $\mathbf{Z}_{L}$:
\begin{equation}
    \mathcal{L}^{\mathrm{CTC}} = -\log P_{\mathrm{CTC}}(\mathbf{y} \mid \mathbf{Z}_{L})
\end{equation}

\subsection{Proposed method}
The overall illustration of our method is shown in Figure~\ref{fig:proposed_model}.
For outputs of some encoder layers, we calculate the probability distribution over vocabularies using Eq. (6) and Eq. (7) to get $\mathbf{Z}_{l}$, which we call an intermediate prediction, and propose to utilize it to condition the prediction of the last layer.
Specifically, we first transform the hidden dimension of $\mathbf{Z}_{l}$ from $\mathcal{|V^{\prime}|}$ to $D$ using a linear layer $\mathrm{Linear}_{\mathcal{|V^{\prime}|} \rightarrow D}(\cdot)$ whose parameters are shared in the model.
Then, it is summed with the original output $\mathbf{X}_{l}^{\mathrm{out}}$ and used as the input of the next layer:
\begin{equation}
    \mathbf{X}_{l+1}^{\mathrm{in}} = \mathrm{LayerNorm}(\mathbf{X}_{l}^{\mathrm{out}}) + \mathrm{Linear}_{\mathcal{|V^{\prime}|} \rightarrow D}(\mathbf{Z}_{l})
\end{equation}
where $\mathrm{LayerNorm}(\cdot)$ denotes the same layer normalization layer as in Eq. (6).
In this way, $\mathbf{X}_{l+1}^{\mathrm{in}}$ contains the information of the intermediate prediction $\mathbf{Z}_{l}$ and we expect the model to leverage this information for the final prediction in Eq. (8) since $\mathbf{Z}_{L}$ is also conditioned on $\mathbf{Z}_{l}$.
We note that each prediction of alignments in the last layer is still conditionally independent.
However, we believe that the performance degradation due to the conditional independence assumption can be alleviated to some extent by conditioning the final prediction on intermediate predictions.

We train a Transformer encoder with intermediate CTC losses in addition to the original CTC loss of the last encoder layer.  
% Intermediate CTC losse is an auxiliary loss, which is calculated for the output of an intermediate layer in the CTC encoder.
Intermediate CTC loss for the $l$-th layer is calculated as follows:
\begin{equation}
    \mathcal{L}_{l}^{\mathrm{inter}} = - \log P_{\mathrm{CTC}}(\mathbf{y} \mid \mathbf{Z}_{l})
\end{equation}
In this study, we take multiple intermediate CTC loss for $K$ intermediate layers and sum those losses:
\begin{equation}
    \mathcal{L}^{\mathrm{inter}} = \frac{1}{K} \sum_{k=1}^{K} \mathcal{L}_{\lfloor\frac{k L}{K+1}\rfloor}^{\mathrm{inter}}
\end{equation}
Unless stated otherwise, we set $L$ and $K$ to 18 and 5 for all experiments, meaning that we take intermediate CTC loss for every 3 layers.
The total training objective is defined as follows:
\begin{equation}
    \mathcal{L} = (1 - \lambda) \mathcal{L}^{\mathrm{CTC}} + \lambda \mathcal{L}^{\mathrm{inter}}
\end{equation}
We set $\lambda$ to 0.5 throughout all experiments.

It is worth noting that if we train a model with Eq. (12) and use Eq. (5) instead of using Eq. (9) as an input of the next layer, it is identical to InterCTC.

\section{Experiments}
To evaluate the effectiveness of our method, we conduct experiments using ESPnet~\cite{watanabe2018espnet}.
In addition to our model, we also evaluate the performances of the autoregressive model, the standard CTC model, Mask-CTC, and InterCTC for comparison.

\subsection{Datasets}
The experiments are conducted on three different ASR corpora: the 81 hours Wall Street Journal (WSJ)~\cite{paul1992design} in English, the 207 hours TEDLIUM2~\cite{rousseau2014enhancing} in English, and the 170 hours AISHELL-1~\cite{bu2017aishell} in Mandarin Chinese.
We use 80-dimensional logmelspec features with 3-dimensional pitch features as inputs and SpecAugment~\cite{park2019specaugment} for data augmentation.
We also apply speed perturbation~\cite{ko2015audio} for TEDLIUM2 and AISHELL-1.
For the tokenization of texts, we use 50 characters for WSJ, 500 subwords created with SentencePiece~\cite{kudo2018sentencepiece} for TEDLIUM2, and 4,231 characters for AISHELL-1.

\begin{table}[th]
\caption{Word error rate (WER) and real time factor (RTF) on TEDLIUM2.
RTF is measured on the dev set.
$\dagger$: 24-layer. Trained with a single intermediate CTC loss and stochastic depth~\cite{huang2016deep}.}
\label{tab:tedlium}
\centering
\begin{tabular}{lcccc}
\hline
model & {dev} & {test} & RTF & Speedup \\
\hline
\multicolumn{5}{l}{\emph{Autoregressive (our work)}} \\
\quad Transformer & 12.6 & 10.2 & 0.290 & 1.00$\times$ \\
\quad + beam search & 10.4 & \textbf{9.2} & 5.596 & 0.05$\times$ \\
\hline
\multicolumn{5}{l}{\emph{Non-autoregressive (previous work)}} \\
\quad Mask CTC~\cite{higuchi2020improved} & - & 10.9 & - & - \\
\quad \quad + DLP~\cite{higuchi2020improved} & - & 10.6 & - & - \\
\quad InterCTC $\dagger$~\cite{lee2021intermediate}  & 10.9 & 10.2 & - & - \\
\quad KERMIT~\cite{fujita2020insertion} & 10.5 & 9.8 & - & - \\
\hline
\multicolumn{5}{l}{\emph{Non-autoregressive (our work)}} \\
\quad CTC & 12.8 & 12.2 & 0.039 & 7.44$\times$ \\
\quad Mask CTC & 11.9 & 10.7 & 0.058 & 5.00$\times$ \\
\quad InterCTC & 10.8 & 10.1 & 0.039 & 7.44$\times$ \\
\quad \textbf{Proposed} & \textbf{9.9} & \textbf{9.4} & 0.041 & 7.07$\times$ \\
\hline
\end{tabular}
\end{table}

\subsection{Experimental Setup}
For hyperparameters of the experiments, we basically follow the autoregressive Transformer architecture as in~\cite{karita2019improving}, which consists of a 12-layer encoder and 6-layer decoder with 4 attention heads and 256 hidden units.
The standard CTC model, InterCTC, and our model consist of an 18-layer encoder to match the number of parameters with the autoregressive model.
We train models for 100 epochs on WSJ and 50 epochs on TEDLIUM2 and AISHELL-1 and average 10 checkpoints with top validation scores to obtain the final model.
We use a batch size of 128 samples for WSJ and TEDLIUM2 and a batch size of 64 samples for AISHELL-1.
Gradients are accumulated~\cite{ott2018scaling} over 2 iterations for all corpora.
The beam width of the autoregressive model is 20 when beam search is used.
Other models are decoded by greedy decoding for CTC.
We don't use any external language model in decoding.
For the model hyperparameters and training configurations of Mask-CTC, we followed the recipe of ESPnet.
Decoding of Mask-CTC is done with 10 decoding iterations.
All training is performed on a single V100 GPU and real time factor (RTF) is measured by decoding with a batch size 1 on a single Intel(R) Xeon(R) Gold 6230 CPU @ 2.10GHz.

\begin{table}[th]
\caption{Word error rate (WER) and real time factor (RTF) on the dev93/eval92 set of WSJ.
RTF is measured on the eval92.
$\dagger$: Same meaning as in Table~\ref{tab:tedlium}.
*: Reference values of RTFs, which could be measured in different environments from ours.}
\label{tab:wsj}
\centering
\begin{tabular}{lcccc}
\hline
model & dev & eval & RTF & Speedup \\
\hline
\multicolumn{5}{l}{\emph{Autoregressive (our work)}} \\
\quad Transformer & 13.5 & 10.5 & 0.574 & 1.00$\times$ \\
\quad + beam search & \textbf{13.2} & \textbf{10.3} & 12.212 & 0.05$\times$ \\
\hline
\multicolumn{5}{l}{\emph{Non-autoregressive (previous work)}} \\
\quad Imputer~\cite{chan2020imputer} & - & 12.7 & - & - \\
\quad InterCTC $\dagger$~\cite{lee2021intermediate}  & 14.9 & 11.8 & - & - \\
\quad Mask CTC~\cite{higuchi2020improved} & 14.9 & 12.0 & 0.063* & - \\
\quad \quad + DLP~\cite{higuchi2020improved} & 13.8 & \textbf{10.8} & 0.074* & - \\
\quad Align-Refine~\cite{chi2020align} & \textbf{13.7} & 11.4 & 0.068* & - \\
\hline
\multicolumn{5}{l}{\emph{Non-autoregressive (our work)}} \\
\quad CTC & 18.7 & 14.9 & 0.036 & 15.94$\times$ \\
\quad Mask CTC & 15.1 & 12.4 & 0.086 & 6.67$\times$ \\
\quad InterCTC & 15.9 & 12.7 & 0.036 & 15.94$\times$ \\
\quad \textbf{Proposed} & \textbf{14.5} & \textbf{11.9} & 0.037 & 15.51$\times$ \\
\hline
\end{tabular}
\end{table}

\begin{table}[th]
\caption{Character error rate (CER) and real time factor (RTF) on AISHELL-1.
RTF is measured on the test set.
$\dagger$: Same meaning as in Table~\ref{tab:tedlium}.}
\label{tab:aishell}
\centering
\begin{tabular}{lcccc}
\hline
model & dev & test & RTF & Speedup \\
\hline
\multicolumn{5}{l}{\emph{Autoregressive (our work)}} \\
\quad Transformer & 5.6 & 6.1 & 0.105 & 1.00$\times$ \\
\quad + beam search & \textbf{4.9} & 5.4 & 1.604 & 0.07$\times$ \\
\hline
\multicolumn{5}{l}{\emph{Non-autoregressive (previous work)}} \\
\quad A-FMLM~\cite{chen2019listen} & 6.2 & 6.7 & - & - \\
\quad LASO-big~\cite{bai2020listen} & 5.8 & 6.4 & - & - \\
\quad ST-NAR~\cite{tian2020spike} & 6.9 & 7.7 & - & - \\
\quad InterCTC $\dagger$~\cite{lee2021intermediate} & 5.2 & 5.5 & - & - \\
\quad KERMIT~\cite{fujita2020insertion} & 6.7 & 7.5 & - & - \\
\quad CASS-NAT~\cite{fan2020cass} & 5.3 & 5.8 & - & - \\
\hline
\multicolumn{5}{l}{\emph{Non-autoregressive (our work)}} \\
\quad CTC & 5.7 & 6.2 & 0.038 & 2.76$\times$ \\
\quad Mask CTC & 5.2 & 5.7 & 0.042 & 2.50$\times$ \\
\quad InterCTC & 5.3 & 5.7 & 0.038 & 2.76$\times$ \\
\quad \textbf{Proposed} & \textbf{4.9} & \textbf{5.3} & 0.050 & 2.10$\times$ \\
\hline
\end{tabular}
\end{table}

\begin{figure*}[th]
\centering
\includegraphics[width=\linewidth]{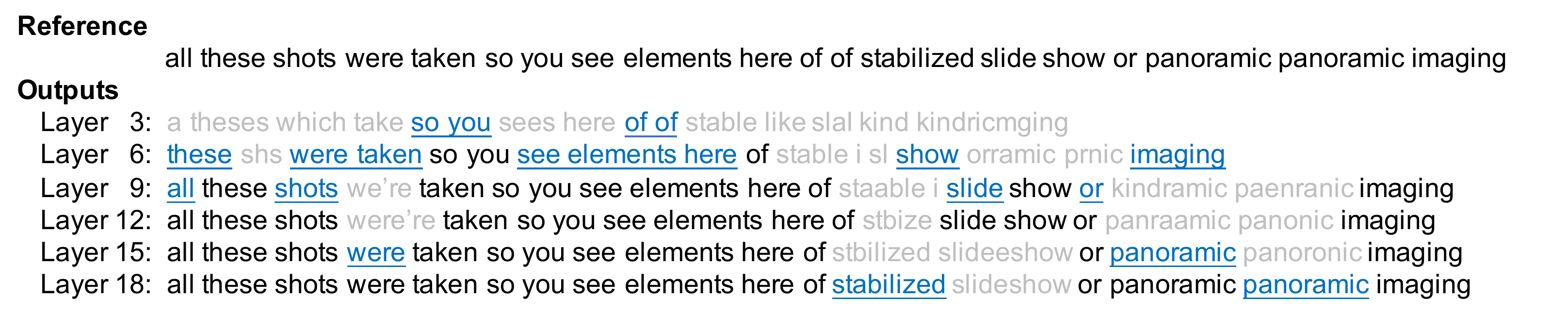}
\caption{An example of outputs in every 3 layers of our 18-layer model on the dev set of TEDLIUM2.
The output is being refined conditioned on the intermediate predictions of previous layers. 
Tokens with errors are colored gray.
Blue underlined tokens are the ones being corrected during decoding.
}
\label{fig:example}
\end{figure*}

\subsection{Results}
Table~\ref{tab:tedlium} shows the results on TEDLIUM2.
Our proposed method gives an improvement over the standard CTC model by 2.9 absolute WER on the dev set and 2.8 absolute WER on the test set with a little deterioration of RTF.
Our model outperforms all previous non-autoregressive works and the autoregressive model without beam search.
Even compared to the autoregressive model with beam search, our model is competitive in terms of WER and the decoding speed is more than 130 times faster.

Table~\ref{tab:wsj} shows the results on WSJ.
Similar to the result of TEDLIUM2, our model achieves better results than the standard CTC model by a large margin (4.2/3.0 absolute WER reduction on the dev93/eval92 set) and decoding speed is much faster than the autoregressive model.
However, different from the result of TEDLIUM2, our model shows worse WER than the autoregressive model, Mask-CTC with dynamic length prediction (DLP) and Align-Refine.
We conjecture that this is because alphabets are used as tokens on WSJ instead of subwords and it is difficult for our model to capture all dependencies between alphabets using only the encoder, while other models enjoy the merit of iterative decoding to model those dependencies.
It is worth noting that our methods can also be applied to the encoder of those other models and could get complementary effects.

Table~\ref{tab:aishell} shows the results on AISHELL-1.
We can see our model performs better than the standard CTC model and previous works of Transformer based non-autoregressive ASR, showing our model is also effective in a language other than English.
Surprisingly, our model also outperforms the autoregressive model with beam search with more than 30 times faster decoding speed.
However, more deterioration of RTF from the standard CTC model is seen compared to the results on WSJ and TEDLIUM2.
This is because we use a relatively larger size of vocabulary (4,231 characters) on AISHELL-1 than other corpora, making the computation time of additional Linear layers and softmax layers in Eq. (7) and (9) longer.

\begin{figure}[th]
    \centering
    \includegraphics[width=\columnwidth]{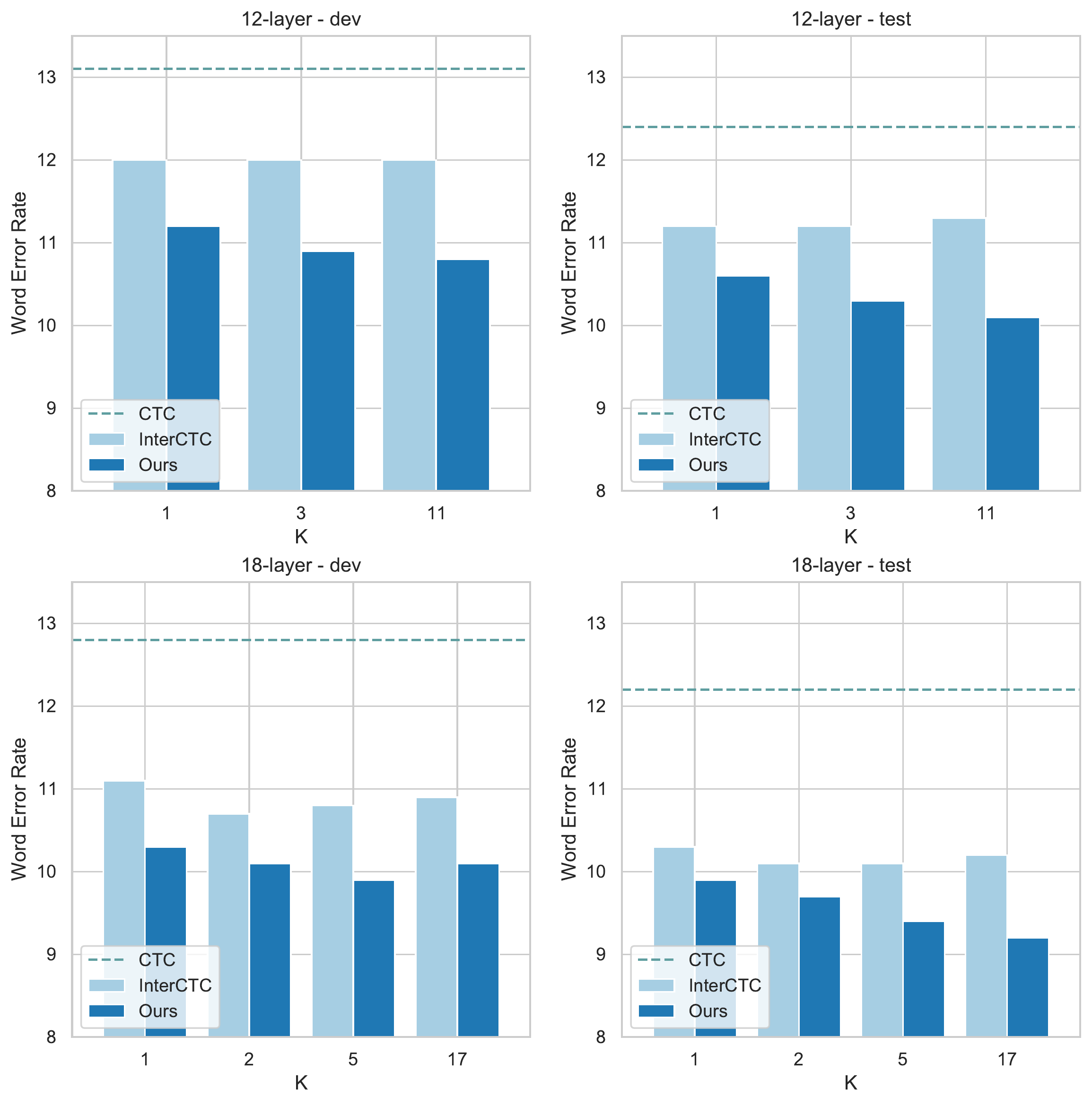}
    \caption{A word error rate (WER) comparison of a different number of intermediate CTC loss $K$ on TEDLIUM2.
    Results of the dev set and test set are shown for 12-layer and 18-layer models.
    Dotted lines indicate the results of the standard CTC model.
    We can see our model can achieve lower WER by increasing $K$, while it is not the case for InterCTC.}
    \label{fig:layercomparison}
\end{figure}

\subsection{Analysis}
Figure~\ref{fig:example} gives an example of outputs on the dev set of TEDLIUM2.
In addition to the output of the last layer (18-th layer), outputs of intermediate layers are shown by taking the argmax of intermediate predictions.
We can see the output is being refined in the encoder layers by conditioning on the intermediate predictions of previous layers.
Since our model refines the output over the frame-level predictions, it can correct insertion and deletion errors in addition to substitution errors.

To investigate the effects of training with multiple intermediate CTC loss and using multiple intermediate predictions, we conduct experiments with a different number of intermediate CTC losses on TEDLIUM2.
We use models of 12-layer and 18-layer to also analyze the effect using a different number of layers.
Results are presented in Figure~\ref{fig:layercomparison}.
For InterCTC, increasing the number of intermediate CTC loss gives no improvement for 12-layer and a little improvement for 18-layer.
This result is similar to the one shown in the original InterCTC paper~\cite{lee2021intermediate}.
On the other hand, different from InterCTC, our model get benefits from increasing the number of intermediate CTC losses well in both 12-layer and 18-layer cases.
We conjecture that this is because our model leverage multiple intermediate predictions to refine the prediction gradually.

\section{Conclusions}
We proposed a method to relax the conditional independence assumption of CTC-based ASR models.
Our method takes CTC losses for the outputs of intermediate layers and utilizes generated intermediate predictions to condition the prediction of the last layer on them.
Experiments on different corpora show our model outperforms a standard CTC model significantly with a little degradation of fast inference speed and achieves comparable performance to a strong autoregressive model with beam search.
Our method is easy to implement and can also be applied to autoregressive models for enhancing encoders, which we leave as future work.

\bibliographystyle{IEEEtran}

\bibliography{mybib}

\end{document}